\documentclass[aps,prd,twocolumn,showpacs,groupedaddress,amsmath,amssymb, superscriptaddress]{revtex4-2}

\usepackage{graphicx}
\usepackage{dcolumn}
\usepackage{bm}
\usepackage{color}
\usepackage[utf8]{inputenc}
\usepackage[bottom]{footmisc}
\usepackage[T1]{fontenc}
\usepackage{mathptmx}
\usepackage{subcaption}
\usepackage[colorlinks,urlcolor=blue,citecolor=blue]{hyperref}
\def\ND{Center for Astrophysics, Department of Physics and Astronomy, University of Notre Dame, Notre Dame, Indiana 46556, USA}
\def\RIT{Center for Computational Relativity and Gravitation, Rochester Institute of Technology, Rochester, New York 14623, USA}
\def\Sogang{Center for Quantum Spacetime, Sogang University, Seoul 04107, Korea}
\def\CRC{Center for Research Computing, University of Notre Dame, Notre Dame, Indiana 46556, USA}
\def\ORNL{National Center for Computational Sciences, Oak Ridge National Laboratory, Oak Ridge, Tennessee 37830, USA}

\begin{document}

\title{Binary neutron star mergers as a probe of quark-hadron crossover equations of state}

\author{Atul Kedia}
\email{atulkedia93@gmail.com}
\affiliation{\ND}
\affiliation{\RIT}

\author{Hee Il Kim}
\email{Corresponding author: khizetta@sogang.ac.kr}
\affiliation{\Sogang}

\author{In-Saeng Suh}
\email{isuh@nd.edu}
\affiliation{\CRC}
\affiliation{\ND}
\affiliation{\ORNL}

\author{Grant J. Mathews}
\email{gmathews@nd.edu}
\affiliation{\ND}

\date{\today}

\begin{abstract}
It is anticipated that the gravitational radiation detected in future gravitational wave (GW) detectors from binary neutron star (NS) mergers can probe the high-density equation of state (EOS). We perform the first simulations of binary NS mergers which adopt various parametrizations of the quark-hadron crossover (QHC) EOS. These are constructed from combinations of a hadronic EOS ($n_{b} < 2~n_0$) and a quark-matter EOS ($n_{b} >~5~n_0$), where $n_{b}$ and $n_0$ are the baryon number density and the nuclear saturation density, respectively. At the crossover densities ($2~ n_0 < n_{b} < 5~ n_0$) the QHC EOSs continuously soften, while remaining stiffer than hadronic and first-order phase transition EOSs, achieving the stiffness of strongly correlated quark matter. This enhanced stiffness leads to significantly longer lifetimes of the postmerger NS than that for a pure hadronic EOS. We find a dual nature of these EOSs such that their maximum chirp GW frequencies $f_{max}$ fall into the category of a soft EOS while the dominant peak frequencies ($f_{peak}$) of the postmerger stage fall in between that of a soft and stiff hadronic EOS. An observation of this kind of dual nature in the characteristic GW frequencies will provide crucial evidence for the existence of strongly interacting quark matter at the crossover densities for QCD.

\end{abstract}

\maketitle

\section{{Introduction}}
Neutron stars (NSs) are an ideal laboratory to probe the properties of matter at very high density. In particular, NS binary systems provide a means to probe the equation of state (EOS) at supranuclear densities (see Refs.~\cite{Baiotti19, Radice20} for general reviews). Indeed, the first detection of gravitational waves (GWs) from the binary NS merger GW170817 by the LIGO-Virgo Collaboration \cite{LIGO-GW170817,LIGO-GW170817eos} has provided fundamental new insights into the nature of dense neutron-star matter \cite{Lattimer12}. Also, measurements of NS masses and radii by the NICER mission give strong constraints on the EOS \cite{Miller19, Riley21, Miller21}.

The tidal effects signaled in the premerger stage are detectable in the ground-based GW observatories \cite{Flanagan08,Hinderer08, Read09tidal}. In the LIGO-Virgo observations, the effective tidal deformability ($\Lambda$) of a NS of mass $M=1.4~\rm{M}_\odot$ was initially deduced to be $\Lambda_{1.4} < 800$ at a 90\% confidence level with a low-spin prior \cite{LIGO-GW170817}. This resulted in a radius constraint for a NS with a mass of $M=1.4 ~\rm{M}_\odot$ to be $R_{1.4} < 13.6 ~\rm{km}$. Subsequently, this was further constrained to be $R_{1.4} = 11.9 \pm 1.4 ~\rm{km}$ \cite{LIGO-GW170817eos}. The deduced primary constraints on the tidal deformability enables a further constraint on the maximum NS mass and the lower limit of the tidal deformability \cite{Annala18,Most18}. Adding the requirement that EOS be consistent with perturbative QCD constrains at densities $> 40~n_0$ \cite{Kurkela10, Annala18, Komoltsev22, Gorda22}, where $n_0$ is the nuclear saturation density, the radius of a maximum-mass NS $R_{\max} < 13.6 ~\rm{km}$ and $\Lambda_{1.4} > 120$ have been reported \cite{Annala18}. It has also been shown that EOSs with a phase transition can give $8.53 ~\rm{km} < R_{1.4} < 13.74 ~\rm{km}$ at the 2~$\sigma$ level and $\Lambda_{1.4} > 35.5$ at a $3~\sigma$ level \cite{Most18}.

A change in the EOS during a phase transition can lead to a variety of dynamical collapse patterns (see Fig. 1 of Ref. \cite{Weih20}). Such changes in the EOS have been identified for the postmerger remnant to produce a noticeable shift of the maximum peak frequency ($f_{peak}$, also known as the $f_2$ frequency in the literature) in the power spectral density (PSD) \cite{Bauswein19, Blacker20, Radice17}. This shift violates the universal relation between $f_{peak}$ and the tidal deformability noted for pure hadronic EOSs \cite{Breschi19}. It is generally expected, however, that the $f_{peak}$ for an EOS with a phase transition will not follow empirical universal relations \cite{Bauswein12a,Hotokezaka13,Bernuzzi14,Rezzolla16,Zappa18}. Hence, observing such a shift could be a decisive indication of the existence of quark matter or other exotic matter at high densities. However, this conclusion is quite model dependent, and some studies have not indicated any significant shift of $f_{max}$ \cite{Most19, Most20}. This shift appears to also depend on how long the merger remnant survives \cite{Weih20,Liebling21,Prakash21}.

The prospect of the postmerger gravitational waves being used to explore the high-density equation of state has been proposed for some time \cite{Bauswein12b}. There have also been a number of recent investigations into the effects of the formation of quark matter in the postmerger \cite{Most20,Bauswein19,Bauswein20,Blacker20,Weih20,Liebling21,Prakash21}. These studies, however, have for the most part investigated effects for a first-order phase transition and the formation of a mixed quark-hadron phase. In all these cases, the first-order transition can soften the EOS. The strength of the order parameter for the high-density phase transition is not known, and it could well be a weakly first-order or simple crossover transition \cite{Pisarski16, Steinheimer11, Hatsuda06, Aoki06, Baym18}. If this is the case, the effect on the postmerger is the opposite. Depending upon the strength of the coupling constants at high density, the pressure from the equation of state can be much larger in the quark matter phase compared to a first-order phase transition. This enhanced stiffness could lead to an altered postmerger in comparison to first-order phase transition EOSs and hadronic EOSs. Thus, by observing an extended postmerger GW signal one could in principle determine not only the order of the transition but the strength of the nonperturbative effects at high density.

We show here that the EOS can be constrained further by the GW signal from the postmerger phase. We examine the behavior of a crossover to the formation of quark-gluon plasma during the postmerger. Though nuclear matter at very high density is asymptotically free, during much of the collapse the quark matter is in the nonperturbative regime of QCD. Here, we show that the properties of quark matter in the nonperturbative regime of QCD formed during the crossover can significantly alter the pressure response of the merged remnant and hence the gravitational radiation during the postmerger phase. We show that the spectra of the gravitational radiation could be used as a sensitive probe of the properties of matter in the crossover to the nonperturbative regime of QCD.

To describe such phase changes in the present work, we utilize quark-hadron crossover (QHC) EOS \cite{Baym19}, henceforth referred to as the QHC19 EOSs (described in detail in Sec. \ref{section:EOS}). As the density increases, it is generally believed that a critical point appears above which a weak first-order chiral transition can occur \cite{Kronfeld12}. Nevertheless, the simplest treatment of the transition from hadronic matter to quark matter is that of a continuous crossover transition without the discontinuous jump associated with a first-order transition. Considering the observations indicating the existence of NSs with high mass ($> 2 ~\rm{M}_\odot$) \cite{Demorest10,Antoniadis13,Fonseca21} and the relatively small maximum radius bounds from the LIGO-Virgo observations, the EOSs that transition from soft to stiff are consistent with these observations and are very interesting for simulations of binary NS mergers.

We study the general dynamics of the mergers and extract the premerger and postmerger characteristics such as the GW frequencies, the tidal deformability, the maximum chirp frequency $f_{max}$, and PSD frequency $f_{peak}$. Our goal is to investigate and identify unique observational signatures of the nature of the QHC EOSs from the binary NS mergers. The postmerger GW emission has been noted to occur in the kilohertz frequency range (1--4 kHz), which is not easily accessible to LIGO/aVirgo/KAGRA and other current GW observatories: however, the third generation GW observatories, the Einstein Telescope and the Cosmic Explorer, will have enhanced sensitivities in this frequency range. Here, we show how observations of characteristic frequencies of the gravitational wave from next-geneneration detectors can identify EOSs with a crossover to quark matter.

The paper is organized as follows. In Sec. \ref{section:EOS} we describe the EOS models we perform merger simulations for. The simulation setup including the initial data and numerical relativity methods are described in Sec. \ref{section:Code}. In Sec. \ref{section:Results}, we discuss the simulation results for the dynamics of premerger and postmerger, followed by the GW frequencies for these phases, and contrast them for several EOSs. Following this, we end with concluding remarks in Sec. \ref{section:Conclusion}.

\section{Equations of state}
\label{section:EOS}

As the baryon density and chemical potential increase, the QCD strong coupling constant $\alpha_s$ approaches unity, and a nonperturbative approach to QCD is imperative. In particular, there is unexplored physics in this region of the quark-matter phase diagram including the generation of constituent quark masses, due to chiral symmetry breaking \cite{Hatsuda94}, and quark pairing leading to color superconductivity \cite{Alford08}.

Various parametrizations of the QHC19 EOS are described in Ref.~\cite{Baym19}. In the original formulation, the low-density hadronic regime $< 2 ~n_0$ of the QHC is described by the Togashi EOS \cite{Togashi13,Togashi17}, which is an extended version of the APR \cite{Akmal98}, and therefore a very soft hadronic EOS. In the present work, we consider other formulations of the hadronic phase as noted below.

The QHC19 EOS accounts for the nonperturbative QCD effects at high densities ($> 5 ~n_0$) in the context of the Nambu–Jona-Lasinio model (see the review in Refs. \cite{Nambu61a,Nambu61b,Buballa05}). Among the four coupling constants, the scalar coupling ($G$), the coefficient of the Kobayashi-Maskawa-'t Hooft vertex ($K$), the coupling for universal quark repulsion ($g_v$), and the diquark strength ($H$), only two ($g_v$ and $H$) are used to construct the model. As these couplings increase, the matter pressure increases \cite{Baym19, Baym18}. For the present work, we utilize three parameter sets given in Ref. \cite{Baym19}, QHC19B [$(g_V,H) = (0.8,1.49)$], QHC19C [$(g_V,H) = (1.0,1.55)$], and QHC19D [$(g_V,H) = (1.2,1.61)$] of QHC19 EOSs. The pressure in the crossover regime ($2 ~n_0 < n < 5 ~n_0$) is described in terms of fifth-order polynomials of the baryonic chemical potential. The tidal deformability of the QHCs satisfies the observational bound from LIGO-Virgo ($\Lambda < 800$ for $M_0 = 1.4 ~\rm{M}_\odot$) \cite{LIGO-GW170817}, where $M_0$ is the gravitational mass at infinite separation of the binary components, and it is similar to that of soft hadronic EOSs such as the SLy and APR4.

\begin{table}[tbpt]
\centering
\caption{Piecewise polytropic parametrization for QHC19 EOSs. The $\rho_5$ and $\rho_6$ densities are given in multiples of $10^{14}$ and $10^{15} ~{\rm{g/cm^{3}}}$, respectively. For all QHC EOSs, $\rho_1$, $\rho_2$, $\rho_3$, and $\rho_4$ are 2.441$\times 10^{7}$, 3.784$\times 10^{11}$, 2.628$\times 10^{12}$, and 1.462$\times 10^{14} ~{\rm{g/cm^{3}}}$ and $\Gamma_1$, $\Gamma_2$, $\Gamma_3$, and $\Gamma_4$ are 1.584, 1.287, 0.622, and 1.357, respectively (identical to the indices by Ref. \cite{Read09}).}
\label{table:1}
\begin{tabular}{@{\extracolsep{8pt}} l c c c c c c @{}}
    \hline \hline
    EOS & $\Gamma_5$ & $\Gamma_6$ & $\Gamma_7$ & $\rho_5$ & $\rho_6$ & Residual \\ \hline
     QHC19B & 2.179 & 3.340 & 2.230 & 2.233 & 1.025 & 0.0081 \\
     QHC19C & 2.382 & 3.479 & 2.191 & 2.699 & 1.105 & 0.0135 \\
     QHC19D & 2.646 & 3.743 & 2.175 & 3.945 & 1.130 & 0.0195 \\ \hline \hline
\end{tabular}
\end{table}
\begin{figure}
    \includegraphics[width=0.5\textwidth]{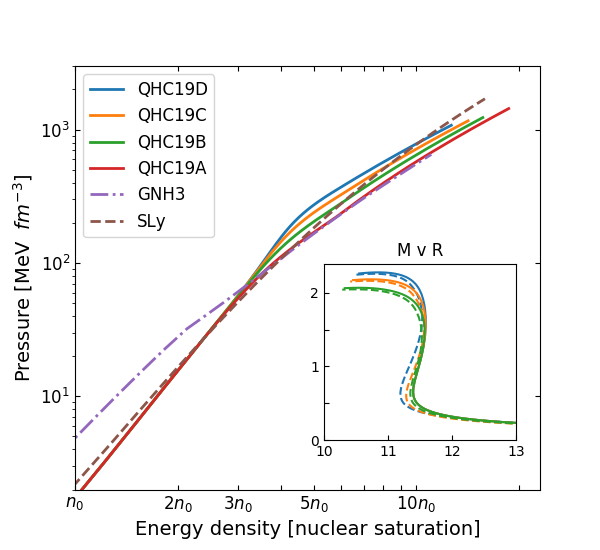}
    \caption{The EOSs studied in this work are shown in $P$ v. $\rho$ (in multiples of nuclear saturation density) here. The inset shows mass-radius ($\rm{M}_\odot$, and km, respectively) relations for raw QHC19 B-D EOSs (solid lines) and their parametrized counterparts (dashed with same color).}
\label{fig:eos}
\end{figure}

We implement the QHC19 EOSs using piecewise-polytropic (PP) fits as described by Ref. \cite{Read09} for our numerical work. We utilize seven polytropic EOS pieces and describe the crust EOS using four fixed pieces among the total seven pieces. However, unlike the original work \cite{Read09}, we do not fix the boundary locations for the highest two density pieces, leaving three high-density $\Gamma$ and two densities $\rho$ undetermined. This allows us to obtain reduced residual values ($< 0.02$) for the QHC EOS fits. The fitting parameters ($\Gamma_5$, $\Gamma_6$, $\Gamma_7$, $\rho_5$, $\rho_6$) thus obtained and their residuals are summarized in Table \ref{table:1}.

Figure \ref{fig:eos} shows the pressure vs energy density plots for the obtained piecewise-polytropic QHC EOSs along with the purely hadronic SLy \cite{SLy} and GNH3 \cite{GNH3} EOSs for reference of soft and stiff hadronic EOSs, respectively. The QHC EOSs exhibit two important defining features as can be seen here. The first is QHCs transition from a soft to a stiff EOS as energy density increases, particularly visible in the crossover density region $\sim 4 n_0$, i.e., the QHCs have a higher pressure in comparison to the hadronic EOSs. The second is that the quark matter phase becomes more stiff as the coupling strengths increase, i.e. going from B to C to D. However, since the original QHC EOSs are interpolated using fifth-order polynomials at the crossover densities, it is not possible to exactly describe the crossover region with a single or a few polytropic pieces as in Ref. \cite{Read09}. Therefore, the mass-radius (M-R) curves (Fig. \ref{fig:eos} inset) show deviations up to a few percent in the maximum NS mass $M_{\max}$ and radius of a NS of maximum mass $R_{\max}$ from those of the original QHCs. The fits have slightly softer pressures near the nuclear saturation densities. Hence, the M-R curves turn around at slightly smaller radii, and the maximum mass is a little bit smaller than that based upon the original QHC19s. The deviations in the radii are up to 3\% in the case of QHC19D, whereas in the case of 19B and C, they are less than 2\%.

Since the maximum mass models involve densities approaching the crossover region, this suggests some uncertainty in these results. Nevertheless, the PP description captures the characteristics of the original QHC EOSs within standard deviations of < 2\%. Although it would be better to utilize the original QHC, we adopted the piecewise-polytropic analytic approximation both for ease of implementation in the \texttt{GRHydro} thorn and for computational speed. Based on this, we conclude that if we made use of the original QHC table, we would obtain simulation results that support our conclusion of a dual nature (to be discussed in Sec. \ref{section:Results}) and the transition from a softer to stiffer EOS even more strongly. From here onward in this work, "QHC EOS" refers to our piecewise-polytropic construction of QHC19 EOSs.

\section{Code description}
\label{section:Code}

We evolve our merger simulations with the use of the numerical relativity software platform, \texttt{the Einstein Toolkit} (ET) \cite{ET}. This is done in full general relativity in three spatial dimensions under the BSSN-NOK formalism \cite{NOK,BSSN1,BSSN2,BSSN3,BSSN4}. We use the \texttt{GRHydro} code \cite{Baiotti,Hawke05,Mosta14} for the general relativistic hydrodynamics based on the Valencia formulation \cite{Valencia1,Valencia2}. We use the HLLE Riemann solver \cite{HLLE}, and the WENO-Z \cite{WENOZ} is used for the fifth-order reconstruction method. Initial data for the NS binary is generated using \texttt{LORENE}\cite{LORENE,LORENE2} for irrotational binaries \cite{Kochanek92,Bildsten92}. The thorn \texttt{Carpet} \cite{Carpet1,Carpet2,Carpet3} is used for the adaptive mesh refinement with six mesh refinement levels and a minimum grid size of 0.3125 in Cactus units ($\approx 461 ~\rm{m}$) for most of the models. The EOSs are supplemented by a thermal pressure component implemented in \texttt{GRHydro} with a constant $\Gamma_{\rm{th}} = 1.8$ \cite{Pietri16}.

The GWs emitted during the evolution are captured using the Newman-Penrose formalism in the form of a multipole expansion of the spin-weighted spherical harmonics of the Weyl scalar $\Psi_4^{(l,m)} (\theta, \phi, t) = \ddot{h}_+^{(l,m)}(\theta, \phi, t) + i \ddot{h}_\times^{(l,m)}(\theta, \phi, t)$. This is then summed over $(l,m)$ modes and integrated twice over time to calculate the $h_+(\theta, \phi, t)$ and $h_\times(\theta, \phi, t)$. The GWs are extracted close to the simulation boundary at $700 ~\rm{M}_\odot$ ($\approx 1033.2 ~\rm{km}$). 

The initial models we evolve in this work have baryonic masses of $M_B =1.45$, $1.50$, $1.55 ~\rm{M}_\odot$ with an initial coordinate separation between centers of $45~\rm{km}$. The corresponding gravitational masses at infinite separation ($M_0$) and the ADM masses ($M_{ADM}$) are summarized in Table \ref{table:2}. Some complementary models are also simulated to confirm our results or to confirm that the successful runs have no numerical artifacts. For the QHC19D, we only ran the case with $M_B=1.55 ~\rm{M}_\odot$ because this did not collapse to a black hole (BH) within the simulation time and leads to the conclusion that lighter masses would not collapse either in the simulation time. This was sufficient to confirm the expected nature based on the results of the QHC19B and QHC19C. As reference hadronic models for a soft and the stiff EOS, we have chosen the SLy and GNH3, respectively.

\section{Results}
\label{section:Results}

We summarize our simulation parameters and outputs in Table \ref{table:2}. The inspiral time is defined as the time at which the maximum density reaches its first minimum. Typically this occurs $ \sim 0.5 ~\rm{ms}$ ahead of the time at which the maximum GW strain occurs. The $t_{BH}$ is the time for a black hole to form from the merger remnant. The time interval $t_{BH}-t_{inspiral}$ defines the duration of the postmerger period of the binaries. Except for the cases of QHC19C with $M_{b}=1.45$, $1.50 ~\rm{M}_\odot$ and QHC19D with $M_{b}=1.55 ~\rm{M}_\odot$, all the other models form a black hole long before $t \simeq 100 ~\rm{ms}$ (the duration of each simulation). The three exceptional cases do not collapse within our simulation time and seemingly end up with the formation of supramassive neutron stars given that their baryonic masses are between $M_{\rm TOV;Baryon}$ and 1.2$\times M_{\rm TOV;Baryon}$\cite{CST92}, where $M_{\rm TOV;Baryon}$ is the maximum baryonic mass of non-rotating NS for each EOS. Their core densities do not cross the $\sim~5~n_0$ density limit for crossover to quark matter, and they simply enter the NS ringdown stage \cite{Bauswein12b}. The SLy binary with $M_B~(M_0) = 1.55 ~(1.40) ~\rm{M}_\odot$ promptly collapses into a black hole. But the QHC19B binary with $M_B~(M_0) = 1.55 ~(1.40) ~\rm{M}_\odot$ sustains a few more dynamical times, while the stiff GNH3 binary with $M_B ~(M_0) =1.55 ~(1.43) ~\rm{M}_\odot$ manifestly shows a delayed collapse into a black hole.

\begin{figure}
    \includegraphics[width=0.52\textwidth]{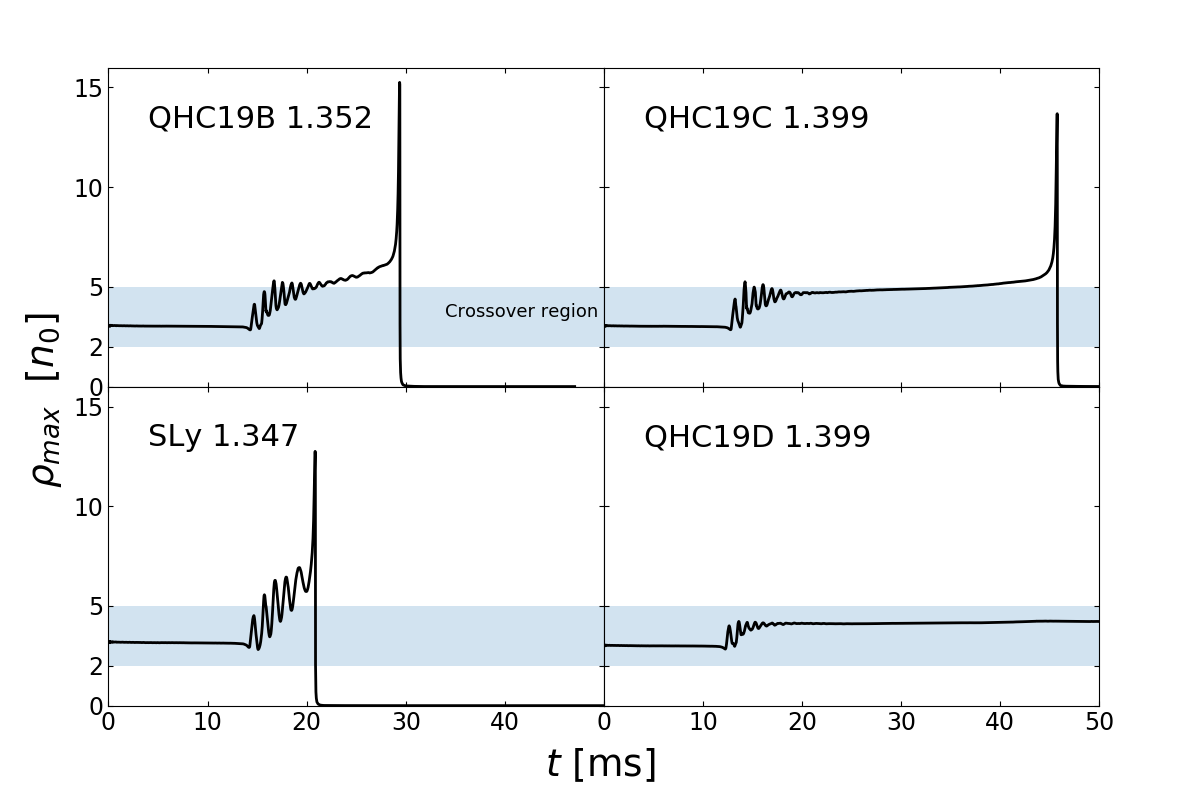}
    \caption{Evolution of maximum rest-mass density vs time for several equations of state employed in the present work. The blue band indicates QHC-crossover densities (note: SLy does not have a crossover: however, indicating the crossover density is suggestive of the need to account for quark matter in hadronic EOSs). In all these cases, the NSs start in the crossover density range (2--5 $n_0$) followed by a rise in density, leading to a collapse to a black hole (in all except the bottom-right panel). The bottom-right case (QHC19D 1.399) does not form a black hole within the simulation time. }
    \label{fig:density}
\end{figure}

\begin{table}[tpb]
\caption{Simulation parameters and evolution outcomes. Units: masses are in ${\rm{M}}_{\odot}$, times in milliseconds, and frequencies in hertz.}
\label{table:2}
\begin{tabular}{@{\extracolsep{4pt}} l c c c c c c c c c c @{}}
    \hline \hline
    EOS & $M_B$ & $M_0$ & $M_{ADM}$ & $t_{inspiral}$ & $t_{BH}$\footnote{An asterisk (*) denotes cases in which a black hole is not formed within the simulation time.} & $f_{peak}$\footnote{A hyphen (-) marks cases for which finding the $f$ modes is challenging due to the PSD plateauing for hundreds of Hz.} & $f_{max}$\\ \hline
    QHC19B & 1.45 & 1.319 & 2.612 & 15.40 & 53.84 & 3150 & 1898\\
    QHC19B & 1.49 & 1.352 & 2.678 & 14.30 & 29.38 & 3291 & 1813\\
    QHC19B & 1.50 & 1.361 & 2.695 & 14.02 & 26.67 & 3336 & 1887\\
    QHC19B & 1.55 & 1.400 & 2.771 & 12.78 & 14.65 &  - & 1796\\ \hline
    QHC19C & 1.45 & 1.319 & 2.612 & 15.46 & * & 3113 & 1864\\
    QHC19C & 1.50 & 1.359 & 2.692 & 14.10 & * & 3200 & 1818\\
    QHC19C & 1.55 & 1.399 & 2.771 & 12.76 & 45.75 & 3287 & 1837\\ \hline
    QHC19D & 1.55 & 1.399 & 2.769 & 12.25 & * & 3183 & 1928\\ \hline
    SLy & 1.45 & 1.323 & 2.620 & 14.97 & 48.05 & 3332 &  1915\\
    SLy & 1.48 & 1.347 & 2.668 & 14.18 & 20.86 & 3545 & 1849\\
    SLy & 1.50 & 1.363 & 2.700 & 13.67 & 16.92 & 3727 & 1913\\
    SLy & 1.55 & 1.404 & 2.779 & 12.49 & 13.31 & - & 1902\\ \hline
    GNH3 & 1.45 & 1.349 & 2.672 & 12.03 & 23.89 & 2534 & 1504 \\
    GNH3 & 1.48 & 1.373 & 2.718 & 11.60 & 20.05 & 2556 & 1557\\
    GNH3 & 1.50 & 1.390 & 2.751 & 11.30 & 19.02 & 2604 & 1525\\
    GNH3 & 1.55 & 1.432 & 2.834 & 10.52 & 14.44 & 2736 & 1595\\
    \hline \hline
\end{tabular}
\end{table}

Figure \ref{fig:density} shows the evolution of the maximum density from our simulations. The densities in the NSs during the merger go well into the crossover range. The NS cores begin in the crossover domain, with central densities of $2.95-3.15 ~n_0$, and throughout the early merger stages stays in it. In the later stages, the maximum density rises either slowly or rapidly, depending on the EOS and the masses of the NSs. Once the maximum density has risen past $\sim 5-6 ~ n_0$ and a few revolutions of the binary have occurred after merger, the density begins to rise rapidly, denoted by the spike, and a collapse occurs.

During the inspiral, the stars in the binary system tidally deform and start to coalesce. As such, $t_{inspiral}$ largely depends on the tidal deformabiliy and the stiffness of the EOSs at densities lower than the initial central densities. As the initial central densities of the QHCs lie in the range of $2.95-3.15 ~n_0$, this is where QHCs are close to but a bit stiffer than the SLy.

The postmerger duration, i.e., the lifetime of the hypermassive neutron star (HMNS), strongly depends on the stiffness at the crossover densities. Below $\sim 3.5 ~n_0$, GNH3 is the stiffest EOS in this study. Hence, the binary mergers based on GNH3 have a longer postmerger duration compared to SLy. On the other hand, the opposite applies for a soft EOS like SLy. This is apparent on Fig. \ref{fig:eos}. SLy is the stiffest at very high densities ($8-20 ~n_0$); however, when this high a density is attained at the core, it is impossible to prevent the merger remnant from collapsing into a black hole due to the high compactness. QHC19 EOSs become stiffer than both SLy and GNH3 at densities $3.5-6 ~n_0$. Because of the increased stiffness, the postmerger remnants from QHC19 binaries have sufficient pressure to avoid the gravitational collapse and exhibit the longest postmerger lifetimes. As the stiffness within QHC models increases, longer lifetimes of their HMNS remnants are directly noticeable. Even the slightly enhanced (stiffened) QHC19B cases produce much longer postmerger duration compared to the hadronic EOSs. The QHC19C with $M_B= ~1.45~\rm{M}_\odot$ and $1.50~\rm{M}_\odot$ does not collapse to a black hole within the simulation time. Only the highest-mass case ($M_B= ~1.55~\rm{M}_\odot$) collapses. In QHC19D, even the highest-mass case does not collapse, and suggests that the lower-mass cases will not either. These longer postmerger periods are all caused by the enhanced stiffness of the QHC19 EOSs at the crossover densities. This nature is in contrast to their similarity with the SLy at densities $< 2.5 ~n_0$ and is the cause of this distinct behavior in the postmerger.

\begin{figure}
    \begin{subfigure}[b]{0.5\textwidth}
        \includegraphics[width=\textwidth]{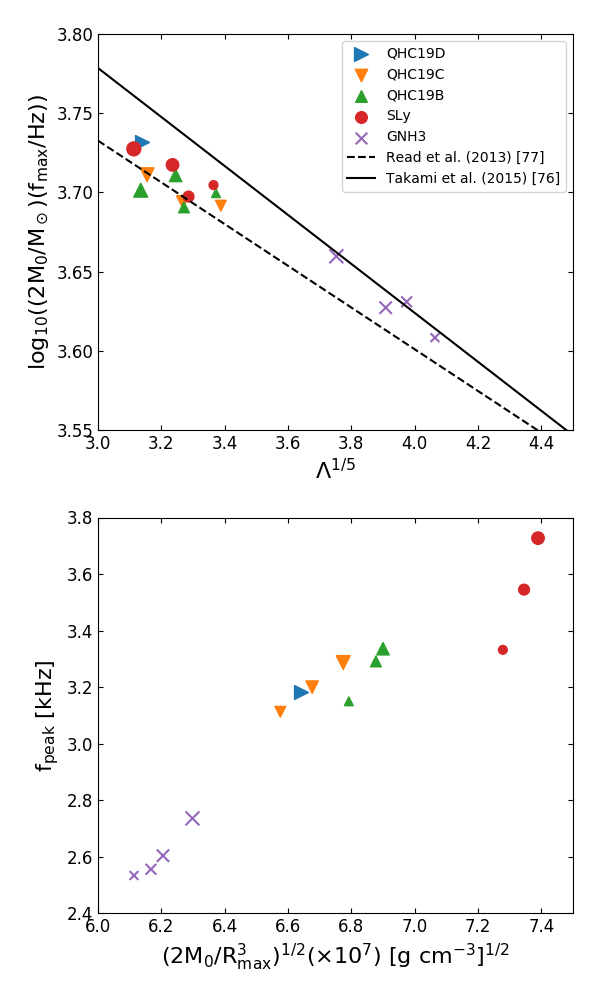}
    \end{subfigure}
    \vspace*{-0.5cm}
    \caption{Top panel shows $f_{max}$ vs the dimensionless tidal deformability ($\Lambda^{1/5}$) along with the universality relations suggested in previous work \cite{Read13, Takami15} as labeled. Lower panel shows $f_{peak}$ vs pseudoaverage rest-mass density $(2M_0/R_{\max}^3)^{1/2}$. For each EOS, increasing sizes of the symbols indicates increase in $M_0$ as listed in Table \ref{table:2}.}
\label{fig:fmax}
\end{figure}

The aforementioned dual nature of the QHC19 EOSs can also be found in analyses of the GW frequencies and the PSD of the strain. Firstly, consider the instantaneous GW frequency of the maximum chirp strain amplitude, $f_{max}=\frac{1}{2\pi}\frac{d\phi}{dt}|_{max}$, where $\phi$ is the phase of the strain (see, e.g., Ref. \cite{Takami15}). It has been suggested that there is a tight universal correlation with the tidal deformability \cite{Takami15,Read13,Bernuzzi14,Bernuzzi15,Rezzolla16,Tsang19,Kiuchi20}. The top panel of Fig. \ref{fig:fmax} shows the relation between $f_{max}$ and the dimensionless tidal deformability ($\Lambda$) for our models along with the universality relations found in Refs.~\cite{Read13, Takami15}. A validation of these universality relations is beyond the scope of this work. However, it should be noted that the $f_{max}$ of the GNH3 cases are closer to the universality curve of Ref. \cite{Takami15}, while that for SLy and QHCs are closer to the universality relation of Ref. \cite{Read13}. Moreover, the $f_{max}$ values for the QHCs are closely aligned with those of the SLy, showing the characteristics of a soft EOS at low densities.

Further, universal relations between $f_{peak}$ and $\Lambda$ have been found in Refs. \cite{Read13,Kiuchi20,Bauswein19,Kiuchi17,Lioutas21} and are well satisfied for pure hadronic EOSs \cite{Breschi19}. The connection between $f_{peak}$ values and other properties, such as the compactness and radius for a fixed fiducial mass, can be also found in Refs. \cite{Foucart16,Lehner16,Bauswein12a,Bauswein12b,Bauswein14,Takami14,Dietrich15,Dietrich17,Maione17}. In the bottom panel of Fig. \ref{fig:fmax}, we show $f_{peak}$ as a function of the pseudoaverage rest-mass density $(2M_0/R_{\max}^3)$ (as was done in in Ref. \cite{Takami15}). Although the pseudoaverage density is not a direct observable, it is shown in Ref. \cite{Bauswein12a} that it can be inferred from $f_{peak}$ and an observation of the mass of the binary. Since the compactness, in general, reflects the stiffness of the EOSs, the stiff GNH3 models are located in the lower-compactness region, while SLy models are in the higher-compactness region. The QHC models are notably clustered in the middle between them. It can be said that the QHCs are mild EOSs, in between soft and stiff EOSs in terms of the compactness. The $f_{peak}$ values of the QHCs are also distinctively clustered in the middle, and they are smaller than those of the SLy models. This further verifies the dual nature of the QHC EOSs; i.e., the QHC EOSs at lower densities behave as soft EOS, as noted from the top panel of Fig. \ref{fig:fmax}, and then at higher densities transition to being stiffer (in comparison to other EOSs) as noted from the bottom panel.

QHC EOSs satisfy the weak empirical trend between $f_{peak}$ and pseudoaverage rest-mass density shown for several hadronic EOSs in Ref. \cite{Bauswein12a, Takami15}. Reference \cite{Takami15} has shown that a tight correlation between the two does not exist. However, an overall trend is present: softer (stiffer) EOSs have a higher (lower) $f_{peak}$ and higher (lower) pseudoaverage rest-mass density. As shown in Fig. \ref{fig:fmax} bottom plot, the QHC EOSs are placed in between the SLy and GNH3. This indicates a stiffened behavior of QHCs in comparison to the hadronic EOSs in the high-density regime. Hadronic EOSs placed on figures like Fig. \ref{fig:fmax} in other works show a small or no shift when moving from the plot of $f_{max}$ to $f_{peak}$. This can be seen in Figs. 11 and 13 of Ref. \cite{Takami15}, in which EOSs are similarly placed in both plots; i.e., higher placed EOSs on the first are also higher placed on the second. QHCs, however, show a significant lateral movement highlighting their dual nature (relatively soft at low densities to relatively stiff at high densities). This observation distinguished QHCs from most EOSs that show the usual behavior.

A recent work has also found a diminished $f_{peak}$ value from their stiffened EOS models \cite{Raithel22}. Since the $\Lambda_{1.35}$ values from QHC EOS are similar to that from SLy, the small $f_{peak}$ of QHCs will violate the universal relation in terms of $\Lambda$ and show a slight shift below the universality relation. The shift is due to the extra stiffening of QHCs in comparison to SLy at crossover densities. This is contrary to the upward shift that appeared for EOSs involving a phase transition (see, Fig. 3 of Ref. \cite{Bauswein19}) due to the softening effect of these models. This opposite effect of the two classes of EOSs may be another crucial method of observation to determine the nature of the nuclear equation of state. However, the $f_{peak}$ values of our SLy models are significantly larger than the previously known values, even slightly violating the universal relation. We have found that this issue comes from the observation that GW frequencies change by $\lesssim$ 0.16 kHz while varying the initial separation by 5 km. As we increase the initial separation from our current value of $45~\rm{km}$, we could get the $f_{peak}$ approaching the universal values. The $f_{peak}$ values of the GNH3 and QHC vary within the error bound described in the last paragraph of this section as the separation changes. We shall address this in further detail in a separate paper.

Recently, another study on the neutron star merger evolution for QHC EOSs has been conducted in Ref. \cite{Huang22}. The EOSs used there are QHC19D and QHC19B, while the hadronic EOS is the Togashi EOS. The study performs mergers for relatively lower-mass cases, so the maximum densities achieved are lower than those shown in Fig. \ref{fig:density}. Their results corroborate ours that a lower $f_{peak}$ is observed for QHC19D when compared with a hadronic EOS. However, they notice a higher $f_{peak}$ for most QHC19B. This discrepancy is due to the higher stiffness of Togashi EOS in comparison to QHC19B at the upper crossover densities ($\sim 3.5~n_0$), whereas in our case, QHC remains stiff all across these densities. Thus, the hadronic EOSs included in this work, the soft SLy EOS and stiff GNH3 EOS, are significantly different from the medium-stiffness Togashi EOS of Ref. \cite{Huang22}. The $f_{peak}$ in their study is slightly smaller compared to our EOS, which can be attributed to us using a lower $\Gamma_{\rm{th}}$, PP QHCs, differences in $f_{peak}$ inference method, and other setup differences, and a detailed comparison study needs to be conducted. However, a study on the dual nature of QHCs from $f_{max}$ and $f_{peak}$, like Fig. \ref{fig:fmax} of this current work, is not conducted in Ref. \cite{Huang22}. Overall, the two works complement each other by analyzing the behavior of the  QHC EOS at different ranges of masses and comparing it with hadronic EOSs of different stiffnesses.

We note that the resolution adopted in this work only corresponds to the medium resolution of Ref. \cite{Pietri16}, which used the same code environment, i.e., the \texttt{GRHydro} and the Carpet thorns in \texttt{the Einstein Toolkit} package. We have performed convergence tests by taking resolutions of 0.375 $\rm{M}_\odot$, 0.3125 $\rm{M}_\odot$, and 0.25 $\rm{M}_\odot$. Owing to the strong dynamical variations and shock formation during the postmerger, the postmerger duration increases as the resolution increases. This is inevitable with the current resolution. However, the characteristic frequencies such as $f_{max}$ and $f_{peak}$ only vary within $0.05 ~\rm{kHz}$, and our qualitative conclusions will not change with increasing resolution. Even with the uncertainty, we anticipate that detecting both $f_{max}$ and $f_{peak}$ from the next-generation GW detectors may reveal effects of the enhanced QCD interactions above the crossover densities.

\section{Conclusion}
\label{section:Conclusion}

We have performed the first simulations of the merger dynamics of binary NSs with QHC19 EOSs and found exciting new features in their dynamical evolution and waveform frequencies. We have shown that neutron stars with QHC EOSs exhibit a dual nature in their evolution pattern. The softness of QHCs in lower densities, $\sim 3 n_0$, is imprinted in their premerger $f_{max}$ frequency, whereas the stiffness in higher densities is imprinted in their postmerger $f_{peak}$ frequency. This dual nature of the QHCs (having both softness and stiffness) can be revealed by the observation of $f_{max}$ and $f_{peak}$ in a GW event. Therefore, in addition to allowing an estimation of $R_{\max}$ or $\Lambda$, NS mergers could reveal (or significantly constrain) quark interaction physics at supranuclear densities. The QHC EOSs were adapted with piecewise-polytropic parametrizations, and they agree to within 2\% error in the M-R relation.

Because of the stiffness of the EOS at the crossover densities, the merger dynamics shows an observably longer postmerger duration compared to that of soft or stiff EOSs. However, it is not easy to quantify the postmerger durations in relation to the dynamical features, since our current numerical setup misses the thermal nuclear EOSs and the microphysics such as neutrino cooling and thermal nuclear interactions. Nevertheless, the hydrodynamical features found in this work will significantly affect the binary NS studies taking into account those realistic considerations. Also, there may be more chances of forming a long-lived NS in the postmerger phase. Moreover, for either equal or unequal mass binaries, the longer lifetime of the core will cause the ejecta dynamics \cite{Metzger20} to show different patterns compared to that of the binary NSs with a normal hadronic EOS. This could affect the electromagnetic counterpart and corresponding nuclear processes in the ejecta. 

Future work should elaborate the results of the current work to give more precise estimates of $f_{max}$ and $f_{peak}$ in conjunction with the universality relations and devise the ways of elucidating the physics of the crossover EOSs. Further, newer QHC EOSs have been formulated recently, adding more stiffness in the crossover and higher densities \cite{Kojo22}. Our findings here suggest that a stronger dual nature would be observed in a study of this stiffer EOS.

\acknowledgements

Work at the Center for Astrophysics of the University of Notre Dame is supported by the U.S. Department of Energy under Nuclear Theory Grant No. DE-FG02-95-ER40934. A.K. acknowledges support from National Science Foundation Grant No. AST-1909534. This research was supported in part by the Notre Dame Center for Research Computing through high performance computing resources. H.I.K. graciously thanks Jinho Kim and Chunglee Kim for continuous support. The work of H.I.K. was supported by Basic Science Research Program through the National Research Foundation of Korea (NRF) funded by the Ministry of Education through the Center for Quantum Spacetime (CQUeST) of Sogang University (Grant No. NRF-2020R1A6A1A03047877). This research used resources of the Oak Ridge Leadership Computing Facility at the Oak Ridge National Laboratory, which is supported by the Office of Science of the U.S. Department of Energy under Contract No. DE-AC05-00OR22725.

Software used was as follows: \texttt{The Einstein Toolkit} (Ref. \cite{ET}; \url{https://einsteintoolkit.org}), \texttt{LORENE} (Refs. \cite{LORENE, LORENE2}), \texttt{PyCactus} (\url{https://bitbucket.org/GravityPR/pycactus}), and \texttt{TOVsolver} (\url{https://github.com/amotornenko/TOVsolver}).

\end{document}